\documentclass[aps,prb,showpacs,groupedaddress,twocolumn]{revtex4}
\usepackage{graphicx}
\usepackage{amsmath}
\usepackage{bm}

\begin{document}

\title{Universal Conductance
Fluctuations in Mesoscopic Systems with Superconducting Leads:
Beyond the Andreev Approximation}

\author {Yanxia Xing$^{1,2}$ and Jian Wang$^{1,*}$}

\address{
$^1$Department of Physics and the center of theoretical and
computational physics, The University of Hong Kong, Hong Kong,
China\\
$^2$Department of Physics, Beijing Institute of Technology, Beijing
100081, China }

\begin{abstract}
We report our investigation of the sample to sample fluctuation in
transport properties of phase coherent normal metal-superconductor
hybrid systems. Extensive numerical simulations were carried out for
quasi-one dimensional and two dimensional systems in both square
lattice (Fermi electron) as well as honeycomb lattice (Dirac
electron). Our results show that when the Fermi energy is within the
superconducting energy gap $\Delta$, the Andreev conductance
fluctuation exhibits a universal value (UCF) which is approximately
two times larger than that in the normal systems. According to the
random matrix theory, the electron-hole degeneracy (ehD) in the
Andreev reflections (AR) plays an important role in classifying UCF.
Our results confirm this. We found that in the diffusive regime
there are two UCF plateaus, one corresponds to the complete
electron-hole symmetry (with ehD) class and the other to
conventional electron-hole conversion (ehD broken). In addition, we
have studied the Andreev conductance distribution and found that for
the fixed average conductance $\langle G\rangle$ the Andreev
conductance distribution is a universal function that depends only
on the ehD. In the localized regime, our results show that ehD
continues to serve as an indicator for different universal classes.
Finally, if normal transport is present, i.e., Fermi energy is
beyond energy gap $\Delta$, the AR is suppressed drastically in the
localized regime by the disorder and the ehD becomes irrelevant. As
a result, the conductance distribution is that same as that of
normal systems.
\end{abstract}


\maketitle

\section{introduction}

It is well known that quantum interference leads to significant
sample-to-sample fluctuations in the conductance at low
temperatures. These fluctuations can be observed in a single sample
as a function of external parameters such as the magnetic field
since the variation of magnetic field has a similar effect on the
interference pattern as the variation in impurity configuration. One
of the fundamental problems of mesoscopic physics is to understand
the statistical distribution of the conductance in disordered
systems\cite{book,Beenakker,add}.

It has been established that in the diffusive regime, the
conductance of any metallic sample fluctuates as a function of
chemical potential, impurity configuration (or magnetic field) with
a universal conductance fluctuation (UCF) that depends only on the
dimensionality and the symmetry of the system.\cite{lee85} The UCF
is given by ${\rm Var} (G/G_0)=2/(16\beta)$, $2/(15\beta)$,
$3/(16\beta)$, $5/(17\beta)$ for quantum dot (QD), quasi-one
dimension (1D), two dimensions (2D) square and three dimensions
cubic sample with $G_0=2e^2/h$. Here the index $\beta$ corresponds
to circular orthogonal ensemble (COE) when the time-reversal and
spin-rotation symmetries are present ($\beta=1$), circular unitary
ensemble (CUE) if time-reversal symmetry is broken ($\beta=2$) and
circular symplectic ensemble (CSE) if the spin-rotation symmetry is
broken while time-reversal symmetry is maintained ($\beta=4$),
respectively.\cite{lee85} In the crossover regime from diffusive to
localized regimes, the conductance distribution was found to be a
universal function that depends only on the average conductance for
quasi-1D, 2D, and QD {\it mesoscopic} systems and
for $\beta=1,2,4$.\cite{saenz2,ucf4} In the localized regime, the
conductance distribution seems to be independent of dimensionality
and ensemble symmetry.\cite{ucf4}

In the presence of a superconducting lead, using random matrix
theory (RMT), the conductance fluctuations in the mesoscopic normal
and superconductor hybrid systems have been studied in the diffusive
regime for quasi-1D
systems\cite{Beenakker,Beenakker1,Beenakker2,Takane} and QD
system.\cite{randmatri} It was found that the UCF in a COE system
shows approximately a twofold increase over the normal systems,
i.e., ${\rm rms}(G_{NS})\simeq2 ~ {\rm
rms}(G_N)$.\cite{Takane,randmatri} Different from the normal
conductor, in the presence of superconducting lead, electron-hole
degeneracy (ehD) plays a similar role of "symmetry". UCF assumes
different value depending on whether ehD is broken or not. According
to RMT,\cite{Beenakker} the Andreev conductance fluctuation ${\rm
rms}(G_{NS})= \sqrt{4.3} ~ {\rm rms}(G_N)$ (with ehD) and ${\rm
rms}(G_{NS})= \sqrt{4} ~ {\rm rms}(G_N)$ (with ehD broken) were
predicted. Up to now, however, most of the investigations on Andreev
conductance fluctuation have been done for systems with ehD
($\epsilon=0$) and low energy regime ($\Delta\ll E_c$ where $E_c$ is
Thouless energy). There is not yet a numerical study on the NS
hybrid system where ehD is broken. In fact, for the existing studies
on the NS hybrid system with ehD, there is no consensus on the
theoretical predicted value of ${\rm rms}(G_{NS})$. Specifically,
concerning the increase factor $\alpha_0$ in ``${\rm
rms}(G_{NS})=\alpha_0 ~ {\rm rms}(G_N)$", a diagrammatic theory
predicted $\sqrt{6}$,\cite{Takane1} and a numerical calculation
using tight binding model gave $\alpha_0=\sqrt{4}$,\cite{Takane} and
the random matrix theory indicated
$\alpha_0=\sqrt{4.3}$\cite{Beenakker} and
$\sqrt{4.5}$.\cite{Beenakker1}

Recently, graphene based normal-metal-superconductor (GNS) systems
were intensively studied because good contacts between the
superconductor electrodes and graphene have been realized
experimentally.\cite{ref16,ref17} In a conventional (quadratic
energy dispersion relation) normal-metal-superconductor (CNS)
system, the usual Andreev reflection (AR) occurs.\cite{ref15} For
GNS systems, AR can be either intravalley or intervalley, which are
called Andreev retroreflection (ARR) and specular Andreev reflection
(SAR), respectively.\cite{ref5} When the excitation energy
$2\epsilon$ is smaller than that of incident energy relative to
Dirac Point $E_F-E_0$, ARR happens, otherwise SAR occurs. At the
transition point $2\epsilon=E_F-E_0$ between ARR and SAR, the
reflection angle $\theta$ (measured relative to the NS junction
normal) jumps from $+90^\circ$ to $-90^\circ$,\cite{Beenakker} the
shot noise vanishes and the Fano factor has a universal
value.\cite{Wangbg} In general, SAR differs from ARR or conventional
AR (CAR) where an extra phase $\pi$ which can be observed in the
quantum interference of the two SAR reflections.\cite{Xing}

So far most of investigations on UCF focus on the Fermi electrons
(quadratic dispersion relation) with the zero or low energy and less
attention is paid on the Dirac electrons. In addition there is no
numerical work reporting Andreev conductance fluctuation when ehD is
broken. It would be interesting to ask the following questions. What
happens to UCF for GNS systems? Is it the same as that in CNS
systems? Is there any difference between ARR and SAR? Which
theoretically predicted value of UCF for the quasi-1D CNS system
(with ehD) is favored? What happened when ehD is broken? What about
the conductance distributions in these systems? It is the purpose of
this paper to address these questions.

In this paper, using the tight-binding model, we carry out a
theoretical study on the sample to sample fluctuation in transport
properties of phase coherent systems with normal
metal-superconductor heterojunction. In view of the possible
difference among CAR, ARR and SAR, we consider both the CNS systems
using the square lattice and GNS system using the honeycomb lattice.
Extensive numerical simulations on quasi-1D and 2D systems in the
presence of a superconducting lead show that when the Fermi energy
is within the superconducting gap $E_F<\Delta$, UCF roughly doubles
the value in the absence of the superconducting lead. This is the
case for both CAR in CNS system and ARR and SAR in GNS system. So
there is no distinct difference between ARR and SAR. Besides,
concerning ehD in the NS hybrid system, new universal classes are
present in agreement with the prediction of RMT.\cite{Beenakker} Two
plateaus of UCF were found in our numerical results, one corresponds
to the complete electron-hole symmetry\cite{ehsymm1} class (with
ehD) and the other to conventional electron-hole conversion (with
ehD broken). It was found that the case of ``ehD broken" decreases
the value of UCF, again in agreement with the theoretical
analysis.\cite{Beenakker} Specifically, in the quasi-1D systems,
${\rm rms}(G_{NS})/{\rm rms}(G_{N})$ for both Fermi and Dirac
electrons is $2.07 \pm 0.04$ that is close to $\sqrt{4.3}$ when ehD
is present while when ehD is broken it is $1.99 \pm 0.08$ that is
close to $\sqrt{4}$. For 2D systems, when ehD is present, ${\rm
rms}(G_{NS})/{\rm rms}(G_{N})$ is $1.91\pm0.07$ for Fermi electrons
and $1.96\pm 0.07$ for Dirac electrons while it is $1.82 \pm 0.08$
when ehD is broken for both Fermi electrons and Dirac electrons.
Furthermore, the different conductance distributions $P(G)$ for the
fixed average conductance $\langle G\rangle$ also indicate this new
symmetry class in localized regime. We also point out that the new
universality class due to the ehD is quite different from the
conventional ensemble symmetries. It was shown numerically that the
conductance distribution $P(G)$ in the deep localized regime for
normal systems is a universal function which depends only on the
average conductance $\langle G\rangle$ but not on the Fermi energies
as well as other parameters.\cite{ucf4} In addition, it does not
seem to depend on the ensemble symmetry and dimensionality of the
system. In the presence of the superconducting lead, our numerical
results for 2D systems with $\beta=1$ show that the conductance
distribution is still an universal function that depends only the
average conductance $\langle G\rangle$. Different from normal
system, however, it depends on whether the system has the ehD.
Finally, when $E_F$ is above $\Delta$, normal transport is present.
We found that the AR is suppressed by the disorder especially in the
localized regime where normal transmission dominates transport
processes. In this case, the ehD is irrelevant and the same
universal conductance distribution is found as that in the normal
systems in the localized regime.

The rest of the paper is organized as follows. In Sec. II, with the
tight-binding representation, the model system including central
disordered region and attached ideal normal lead and superconducting
lead is introduced. The formalisms for calculating the conductance
and fluctuation of conductance are then derived. Sec. III gives
numerical results along with detailed discussions. Finally, a brief
summary is presented in Sec. IV.

\section{model and Hamiltonian}

\begin{figure}
\includegraphics[bb=9mm 12mm 199mm 109mm,
width=7.5cm,totalheight=3.8cm, clip=]{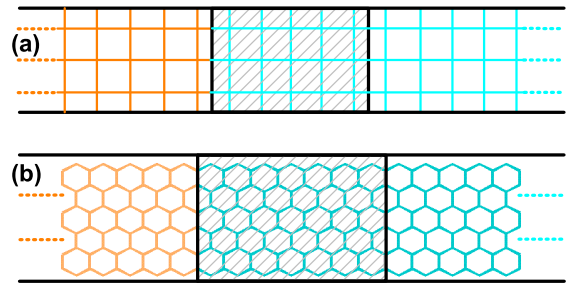} \caption{
(Color online) Schetch of CNS [panel (a)] and GNS [panel (b)]
system, in which ideal superconducting lead (left, orange), normal
lead (right, blue) and disordered normal scattering region (shadowed
blue region) are concluded.} \label{structure}
\end{figure}

The scattering theory of electronic conduction is developed by
Landauer,\cite{Landauer} Imry,\cite{Imry} and
B$\ddot{u}$ttiker.\cite{Buttiker} It provides a complete description
of quantum transport in the system without electron-electron
interactions. A mesoscopic conductor can be modeled by a
phase-coherent disordered region connected by ideal leads (without
disorder) to two electron reservoirs (normal metal or
superconductor), which are in equilibrium at zero temperature with
fixed electrochemical potential (or Fermi energy) $E_F$. Here we
assume that the central scattering region is normal region, the same
as the right normal lead. Then the total system Hamiltonian
\begin{equation}
 H = H_{S} + H_{N} + H_{T}\label{H}
 \end{equation}
where $H_{S}$, $H_{N}$ and $H_{T}$ are the Hamiltonian of
superconducting lead (orange region in Fig.1), semi-infinite normal
ribbon (blue region in Fig.1) and tunneling between the normal
region and superconducting terminal, respectively.

Two kinds of structure were considered in this paper: the structure
with quadratic energy dispersion [square lattice, Fig.1(a)] and
structure with conical energy spectrum [honeycomb lattice,
Fig.1(b)]. In the absence of the superconductor, the whole system
$H_0$ including $H_S$, $H_N$ and $H_C$ can be written in the
tight-binding representation:\cite{ref19,ref20}
\begin{eqnarray}
H_{0} = \sum_{\bf i} (E_0+\delta\epsilon_{\bf i}) a^{\dagger}_{\bf
i} a_{\bf i}
      -\sum_{<{\bf ij}>} t  a_{\bf i}^{\dagger} a_{\bf
      j}\label{H0}
\end{eqnarray}
where ${\bf i}=(i_x , i_y)$ is the index of the discrete square
lattice or honeycomb lattice site which is arranged as in inset of
Fig.1. Here $a_{\bf i}$ and $a_{\bf i}^{\dagger}$ are the
annihilation and creation operators at the discrete site ${\bf i}$.
$E_0$ is the constant on-site energy. In the square lattice, $E_0$
is center of energy band, and in honeycomb lattice, $E_0$ is the
energy reference point (the Dirac Point). $\delta\epsilon_{\bf i}$
is random on-site potential which is nonzero only in the center
region to simulate the disordered scattering region. Here
$\delta\epsilon_{\bf i}$ is uniformly distributed with
$\delta\epsilon_{\bf i}=[-w/2,w/2]$ where $w$ is disorder strength.
The data for fluctuations are obtained by averaging over up to
10,000 disorder configurations and the data for distribution are
obtained over 1,000,000 disorder configurations. The second term in
Eq.(2) is the nearest neighbor hopping with hopping elements $t$ and
``$<>$" denotes the sum over the nearest sites.

Due to the superconductor, it is convenient to write the Hamiltonian
$H$ in the Nambu representation,\cite{Nambu}. In this representation
the Fermi energy of the right normal lead in equilibrium (at zero
bias) is set to be the superconductor condensate. It is
conventionally set to zero. As a result, the spin up electrons and
the spin down holes have the positive and negative energy,
respectively. Taking this into account the Hamiltonian (\ref{H0}) is
cross multiplied by spin representation. $H_N$ and $H_T$ in
Eq.(\ref{H}) can be rewritten as $H_{0,N/T}\otimes \sigma_z$ and
\begin{equation}
 H_{S}= \left(
\begin{array} {cc}
H_{0,S} & \tilde{\Delta} \\
\tilde{\Delta}^* & -H_{0,S}
\end{array}
\right)
\end{equation}
where $\tilde{\Delta}=\Delta e^{i\varphi}$ is the energy gap or the
pair potential of the semi-infinite superconducting lead. Here we
can assume $\tilde{\Delta}=\Delta$ to be a real parameter by
selecting a special phase of the superconductor lead in our
calculation.\cite{realGap}

In the calculation, for simplicity we set external voltage in the
normal and superconducting terminal as $V_N=V$, $V_S=0$. The current
flowing from the normal lead can be calculated from the
Landauer-B$\ddot{u}$ttiker formula:\cite{footnote}
\begin{eqnarray}
&&J_{N}=J^e_{N}-J^h_{N}\nonumber \\
&&J^{e/h}_{N}=\pm\frac{e}{\hbar}\int\frac{dE}{2\pi}~
\left\{T_{e/h}(E)[f_\pm(E)-f_0(E)]\right .\nonumber \\
&&~~~~~~~~~~~~~~~~~~~~~~\left.T_A(E)[f_{\pm}(E)-f_{\mp}(E)]\right\}\label{Landuer}
\end{eqnarray}
where $e$ is the electron charge,
$f_0(E)=\left[e^{E/k_B\mathcal{T}}+1\right]^{-1}$ is the Fermi
distribution in the superconducting lead, $f_{\pm}(E)=\left[e^{(E\mp
eV_N)/k_B\mathcal{T}}+1\right]^{-1}$ are the Fermi distribution
functions in the normal terminal for the electrons and holes,
respectively. $T_{e/h}$ is the transmission coefficient that the
particles incident from superconducting lead traverse to the normal
terminal as electrons/holes and $T_A$ is AR coefficient representing
the reflection probability that the incident electrons from the
normal terminal are reflected as holes or vice versa. Note that the
two processes are symmetric and have the same AR coefficient $T_A$.
$T_{e/h}$ and $T_A$ are calculated from
\begin{eqnarray}
&&T_{e}={\rm Tr}\{\Gamma^N_{\uparrow\uparrow}[G^r\Gamma^S
G^a]_{\uparrow\uparrow}\},~~ T_{h}={\rm
Tr}\{\Gamma^N_{\downarrow\downarrow}[G^r\Gamma^S
G^a]_{\downarrow\downarrow}]\}\nonumber \\
&&T_A={\rm Tr}[\Gamma^N_{\uparrow\uparrow}G^r_{\uparrow\downarrow}
\Gamma^N_{\downarrow\downarrow}G^a_{\downarrow\uparrow}] ={\rm
Tr}[\Gamma^N_{\downarrow\downarrow}G^r_{\downarrow\uparrow}
\Gamma^N_{\uparrow\uparrow}G^a_{\uparrow\downarrow}]
\end{eqnarray}
the line-width function
$\Gamma^{N/S}(E)=i[\Sigma_{N/S}^r(E)-\Sigma_{N/S}^{r\dagger}(E)]$.
The Green's function
$G^r(E)=[G^a(E)]^{\dagger}=[EI-H_{C}-\Sigma^r_{N}(E)-\Sigma^r_{S}(E)]^{-1}$
where $H_{C}$ is Hamiltonian matrix of the central scattering region
and $I$ is the unit matrix with the same dimension as that of
$H_{C}$, $\Sigma_{l=N,S}^r$ is the matrix of retarded self-energy
from the normal/superconducting lead with the only nonzero elements
in the sub-block that are neighbor of normal or superconducting
lead. The self-energy is calculating according to
$\Sigma^r_{l}=H_{Cl}g^rH_{lC}$ where $H_{Cl}$ ($H_{lC}$) is the
coupling from central region (leads) to leads (central region) and
$g^r$ is the surface retarded Green's function of semi-infinite lead
which can be calculated using a transfer matrix
method.\cite{transfer} Due to electron-hole symmetry,
$T_e(E)=T_h(-E)$ and $T_A(E)=T_A(-E)$, which leads to
$J_{N}=2J^e_{N}=-2J^h_{N}$.

At zero temperature limit, the energy dependent conductance can be
expressed as:
\begin{eqnarray}
 G_{NS}(E_F)&=&d(J^e_N-J^h_N)/dV \nonumber \\
 &=&\frac{e^2}{h}
 \left\{[T_{e}(E_F)+T_{h}(-E_F)]+4T_{A}(E_F)\right\}\nonumber \\
 &=&\frac{2e^2}{h}
 \left[T_{e/h}(\pm E_F)+2T_{A}(E_F)\right].\label{conduct}
\end{eqnarray}
When the incident energy $E_F<\Delta$, there is no normal
quasi-particle transport $T_{e/h}=0$ and only AR contributes to
conductance $G$. We will focus mainly on this quantity in this
paper. In this case, the conductance fluctuation defined as ${\rm
rms}(G)=\sqrt{\langle [G-\langle G\rangle]^2\rangle}$ becomes
\begin{eqnarray}
{\rm rms}(G)=\frac{4e^2}{h}\sqrt{\langle T_A^2\rangle-\langle
T_A\rangle^2}
\end{eqnarray}
where $\langle ... \rangle$ denotes averaging over an ensemble of
samples with different disorder configurations of the same strength
$w$. When $E_F$ is beyond superconducting energy gap $\Delta$,
normal transmission $T_{e/h}$ is present, conductance variance now
consists of three components: (1). the Andreev related fluctuation
${\rm Var}(G)_{\rm Andr}$ from AR coefficient $T_{A}$. (2). the
normal fluctuation ${\rm Var}(G)_{\rm Norm}$ from normal
transmission coefficient $T_{e/h}$. (3). the cross term ${\rm
Var}(G)_{\rm cross}$. They are expressed as
\begin{eqnarray}
&& {\rm Var}(G)={\rm Var}(G)_{\rm Andr}+{\rm Var}(G)_{\rm Norm}+{\rm
Var}(G)_{\rm cross}\nonumber \\
&&=[\frac{e^2}{h}]^2\left[\left\langle (4\delta
T_A)^2\right\rangle+\left\langle (\delta
T_N)^2\right\rangle+8\left\langle \delta T_A\delta
T_N\right\rangle\right]\label{Var}
\end{eqnarray}
where $\delta T_A=T_A-\langle T_A\rangle$, $\delta
T_N=(T_e+T_h)-\langle T_e+T_h\rangle$.

\section{results and discussion}

In the numerical calculations, the energy is measured in the unit of
the nearest coupling elements $t$. For the square lattice,
$t=\frac{\hbar^2}{2m^* a^2}$ with $m^*$ the effective electron mass
and $a$ the lattice constant. For the honeycomb lattice,
$t=\frac{2}{3b}\hbar v_F$ with the carbon-carbon distance
$b=0.142nm$ and the Fermi velocity $v_F=0.89\times10^6ms^{-1}$. The
size of the scattering region $N \times M$ is described by integer
$N$ and $M$ corresponding to the width and length, respectively. For
example in Fig.1, the width $W=N a$ with $N=3$, the length $L=M a$
with $M=5$ in the panel (a), and the width $W=N \times 3b$ with
$N=3$, the length $L=M\times \sqrt{3}b$ with $M=7$ in the panel (b).

\begin{table}
\caption{The parameter $E_F$, $E_0$, $\Delta$ used in the square
lattice model and honeycomb model. The different columns are
corresponding to the different transport processes denoted by `AR',
`ARR', `SAR' ,`NT' and so on. Here, `AR' is for the pure
conventional AR assisted tunneling processes (only conventional AR
exists) in square lattice. `ARR` and `SAR' denotes the pure ARR
assisted process and pure SAR assisted process in honeycomb lattice,
respectively. `NT' is for the transport beyond the superconducting
Gap where NT can also contribute to the transport processes.}
\begin{tabular}{c c c c | c c c c | c c c c} \hline\hline sq \\ \hline
AR & $E_F$ & $E_0$ & ~$\Delta$~ & AR & $E_F$ & ~$E_0$~ & ~$\Delta$~
& NT & $E_F$ & ~$E_0$~ & ~$\Delta$~
\\
1 & 0 & 2.1 & 0.1 & 4 & 0.2 & 2.2 & 0.3 & 1 & 0.2 & 2.2 & 0.3
\\
2 & 0 & 2.3 & 0.1 & 5 & 0.3 & 2.3 & 0.4 & 2 & 0.3 & 2.3 & 0.4
\\
3 & 0 & 2.4 & 0.1 & 6 & 0.4 & 2.4 & 0.5 & 3 & 0.4 & 2.4 & 0.5
\\ \hline hc \\ \hline
ARR & $E_F$ & ~$E_0$~ & ~$\Delta$~ & SAR & $E_F$ & ~$E_0$~ &
~$\Delta$~ & ~NT~ & $E_F$ & ~$E_0$~ & ~$\Delta$~
\\
1 & 0   & 0.6 & 0.1 & 1 & 0.6 & 0.0 & 0.7 & 1 & 0.7 & 0   & 0.5
\\
2 & 0   & 0.7 & 0.1 & 2 & 0.6 & 0.1 & 0.7 & 2 & 0.7 & 0.1 & 0.5
\\
3 & 0   & 0.8 & 0.1 & 3 & 0.7 & 0.0 & 0.8 & 3 & 0.7 & 0   & 0.3
\\
4 & 0   & 0.9 & 0.1 & 4 & 0.7 & 0.1 & 0.8 & 4 & 0.7 & 0.1 & 0.3
\\
5 & 0.1 & 0.7 & 0.2 &   &     &     &     & 5 & 0.7 & 0   & 0.1
\\
6 & 0.1 & 0.8 & 0.2 &   &     &     &     & 6 & 0.7 & 0.1 & 0.1
\\ \hline\hline
\end{tabular}
\label{parameter}
\end{table}

As documented in the literature, in order to get the saturated UCF
plateaus,\cite{saenz2,ucf4,LiDafang} the number of transmission
channels for incoming electron should be large enough in the
numerical calculation. We denote $N_c$ as the chain number which
determines directly the number of channels. $N_c$ is defined in the
following way: take Fig.1 as an example, in panel (a), $N_c=3$ and
$N_c=6$ in panel (b). In 2D systems we set $N_c=40$ and $60$. For
quasi 1D systems we use only $N_c=40$ because it is more
computational demanding than 2D systems. To get a larger channel
number, the incident energy $E_F$ should be set away from the bottom
of energy band $E_b$. In the square lattice, to mimic the parabolic
energy spectrum for Fermi electrons, the constraints for incident
energy $E_F<E_b+2t$ and Andreev reflected energy $-E_F<E_b+2t$ are
needed. While for Dirac electrons in the honeycomb lattice, the
absolute value of relative incident energy (to the Dirac point
$E_0$) $|E_F-E_0|<t$ and relative Andreev reflected energy
$|-E_F-E_0|<t$ are set.

In Table.\ref{parameter}, we list all the parameters used in the
following calculations including the incident energy $E_F$, the
superconducting gap $\Delta$ and the on-site energy $E_0$ which is
the center of energy band for the square lattice and the Dirac Point
for the honeycomb lattice. From these parameters in the clean system
with NS heterojunction, we can easily calculate the channel number
for electron or hole, AR coefficient $T_A$ and the normal
transmission coefficient of electron or hole $T_{se/sh}$ for Fermi
energy beyond superconducting gap. At the same time, we can also get
the normal transmission coefficient $T_{ne/nh}$ in normal system
without the superconducting lead.

\subsection{Conductance fluctuation and conductance distribution in the diffusive regime}

\begin{figure}
\includegraphics[bb=11mm 10mm 186mm 140mm,
width=8.5cm,totalheight=6.2cm, clip=]{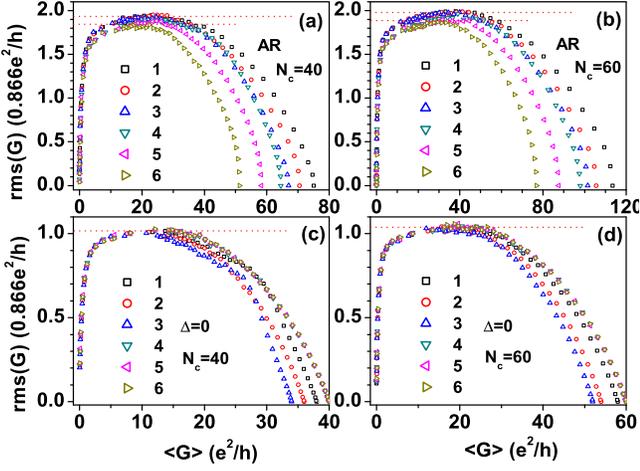} \caption{
(Color online) panel(a) and panel(b): in the presence of
superconducting lead, conductance fluctuation ${\rm rms}(G)$ vs
average conductance $\langle G\rangle$ in the square lattice for
$N_c=40$ and $N_c=60$, respectively. The symbol is labeled in first
column in Tab.\ref{parameter} for the square lattice case denoted by
`sq'. The red dotted lines indicate two plateaus in the values (with
the unit of $0.866e^2/h$) of $1.94\pm0.03$ and $1.85\pm0.05$ in
panel (a) and $1.98\pm0.02$ and $1.89\pm0.03$ in panel (b). For
comparison, corresponding to panel (a) and panel (b), in panel (c)
and (d), we plot ${\rm rms}(G)$ vs $\langle G\rangle$ in the absence
of superconducting lead, i.e., $\Delta=0$, respectively. The
plateaus in the values of $1.02\pm0.02$ and $1.04\pm0.02$ in the
unit of $0.866e^2/h$ are indicated in panel (c) and panel (d). The
system size: $N_c=40$ corresponds to width $W=40a$, considering the
square shape sample, we set $L=40a$. For $N_c=60$, we have $W=60a$,
$L=60a$.} \label{square2D}
\end{figure}

\begin{figure}
\includegraphics[bb=11mm 9mm 197mm 226mm,
width=6cm,totalheight=7cm, clip=]{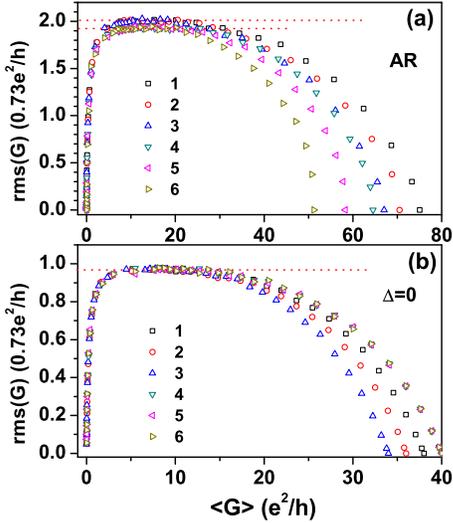} \caption{
(Color online) Same as Fig.\ref{square2D} except the model is
quasi-1D square lattice with chain number $N_c=40$. The system size:
width $W=40a$, length $L=1000a$. In the unit of $0.73e^2/h$, two
plateaus with the values of $2.01\pm0.02$ and $1.93\pm0.03$ in panel
(a) and a single plateau in the value of $0.97\pm0.01$ is indicated
in panel (b).} \label{square1D}
\end{figure}

\begin{figure}
\includegraphics[bb=11mm 10mm 186mm 140mm,
width=8.5cm,totalheight=6.2cm, clip=]{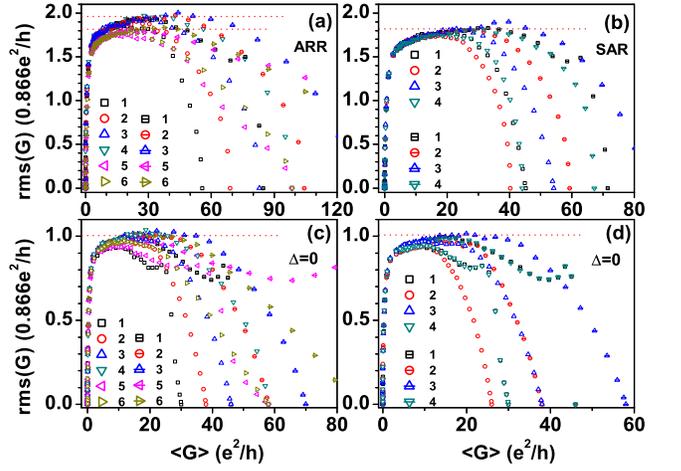} \caption{
(Color online) ${\rm rms}(G)$ contributed by ARR [panel(a)] and SAR
[panel(b)] vs $\langle G\rangle$ in 2D honeycomb lattice for
$N_c=40$ [open symbols] and $N_c=60$ [symbols with `-']. The symbols
are labeled in second column in Tab.\ref{parameter} for the
honeycomb lattice case denoted by `hc'. The red dotted lines
indicate two plateaus with the values of $1.96\pm0.03$ and
$1.82\pm0.02$ in the unit of $0.866e^2/h$ in panel (a) and a single
plateau with the value of $1.82\pm0.02$ in panel (b). Panel (c) and
(d): ${\rm rms}(G)$ contributed by normal quasi-particle
transmission $T_{ne}$ vs $\langle G\rangle$ in case of $\Delta=0$,
corresponding to panel (a) and (b), respectively. The plateaus in
the values of $1.00\pm0.02$ in the unit of $0.866e^2/h$ are
indicated in panel (c) and panel (d). The system size: for $N_c=40$
its width is equal to $W=60b$; considering the square shape sample,
the length $L=35 \sqrt{3}b$. Similarly, for $N_c=40$ its width is
$W=90b$, $L=52 \sqrt{3}b$.} \label{honey2D}
\end{figure}

\begin{figure}
\includegraphics[bb=11mm 10mm 188mm 141mm,
width=8.5cm,totalheight=6.2cm, clip=]{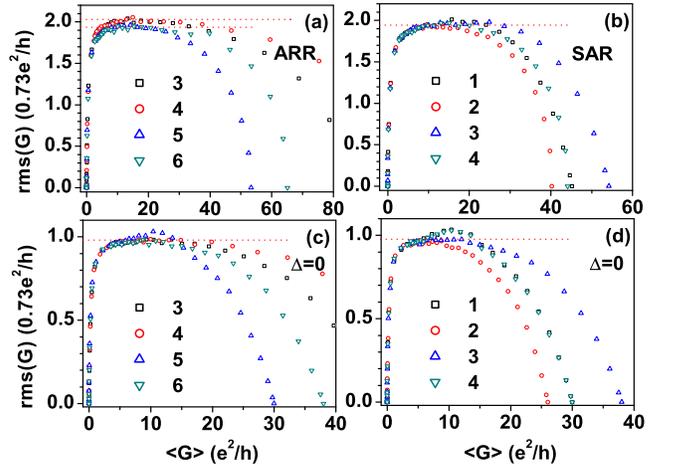} \caption{
(Color online) Same as Fig.\ref{square2D} except the model is
quasi-1D honeycomb lattice with chain number $N_c=40$. The system
size: width $W=60b$, length $L=500 \sqrt{3}b$. The red dotted lines
indicate two plateaus with the values of $2.03\pm0.03$ and
$1.93\pm0.03$ in the unit of $0.73e^2/h$ in panel (a), a single
plateau with the value of $1.94\pm0.04$ in panel (b), the value of
$0.98\pm0.02$ in panel (c) and panel (d).} \label{honey1D}
\end{figure}

We first examine conductance fluctuations in the diffusive regime.
In our calculation the size of 2D square lattice is set to be $40
\times 40$ for $N_c=40$ and $60\times 60$ for $N_c=60$. The size of
2D honeycomb lattice is chosen to be $20\times 35$ for $N_c=40$ and
$30\times 52$ for $N_c=60$. For quasi-1D systems, the size is chosen
to be $40 \times 1000$ in square lattice and $20\times 500$ in
honeycomb lattice with $N_c=40$. In Fig.\ref{square2D} and
\ref{square1D}, \ref{honey2D} and \ref{honey1D}, we plot conductance
fluctuations ${\rm rms}(G)$ vs the average conductance $\langle
G\rangle$ in 2D square lattice, quasi-1D square lattice, 2D
honeycomb lattice and quasi-1D honeycomb lattice, respectively. Each
point in the figure is obtained by averaging over 10,000
configurations. Different parameters used in all figures are
tabulated in Table.\ref{parameter}.

From Fig.\ref{square2D}-\ref{honey1D}, we see following general
behaviors. (1). in the localized regime where $\langle G\rangle<1$,
all the curves collapse into a single curve indicating the universal
behavior of the conductance distribution function.\cite{ucf4} (2).
in the diffusive regime where $\langle G\rangle>1$, there is a
plateau region for ${\rm rms}(G)$ where the fluctuation is nearly
independent of average conductance $\langle G\rangle$ and other
system parameters. This is the regime for the universal conductance
fluctuation. The plateau value [labeled by red dotted line in top
panels] is approximately twice the value of the known UCF values
${\rm rms}(G)=0.866e^2/h$ for 2D system and $0.73e^2/h$ for quasi-1D
system [labeled by red dotted line in bottom panels]. This doubling
seems to be true for both Fermi electrons (square lattice)
[Fig.\ref{square2D} and Fig.\ref{square1D}] and Dirac electrons
(graphene system) [Fig.\ref{honey2D} and Fig.\ref{honey1D}]. (3).
There are two separate UCF plateaus for the AR assisted transport
processes in the CNS system [Fig.\ref{square2D}(a),
Fig.\ref{square1D}(a)] and the ARR assisted transport processes in
GNS system [Fig.\ref{honey2D}(a),Fig.\ref{honey1D}(a)], while for
the SAR assisted transport processes [panel (b)] there is only one
UCF plateau. It appears that this difference in UCF can be used to
distinguish ARR and SAR. However, it turns out to be incorrect when
considering the ehD ``symmetry". In Fig.\ref{honey2D}(a) and
Fig.\ref{honey1D}(a), Andreev conductance fluctuations corresponding
to $E_F=0$ (with ehD) and $E_F\neq 0$ (ehD broken) from diffusive
regime all the way to localized regime are plotted and two UCF
plateaus associated to ehD ``symmetry" are then indicated. For SAR
in graphene systems, we have $|E_F|>|E_0|\ne 0$ (ehD broken).
Fig.\ref{honey2D}(b) and Fig.\ref{honey1D}(b) then show only one UCF
plateau. (4). Denoting the increase factor $\alpha_0$ through the
relation ${\rm rms}(G_{NS})=\alpha_0 ~ {\rm rms}(G_N)$ [$G_N$ is
shown in bottom panels in Fig.\ref{square2D}-\ref{honey1D}] in the
plateau region in diffusive regime, it is very different for square
2D system and quasi-1D system and slightly different for Fermion
electrons and Dirac electrons. Specifically, our results for the
quasi-1D systems for both Fermi electrons and Dirac electron is as
follows: (a). when ehD is present ${\rm rms}(G_{NS})/{\rm
rms}(G_{N})$ is $2.07\pm0.04$ that is very close to $\sqrt{4.3}$.
(b). when ehD is broken it is $1.99 \pm 0.08$ that is close to
$\sqrt{4}$. For 2D systems, when ehD is present, ${\rm
rms}(G_{NS})/{\rm rms}(G_{N})$ is $1.91\pm0.07$ for Fermi electrons
and $1.96\pm0.07$ for Dirac electrons. When ehD is broken it is
$1.82\pm0.08$ for both Fermi electrons and Dirac electrons. (5). For
larger $\langle G\rangle$ (in the ballistic regime) the conductance
fluctuation falls down quickly to zero. This is because the number
of conducting channels $N_c$ is finite\cite{saenz2,LiDafang,ucf4}.
The width of plateau region is longer with a larger $N_c$. In the
limit of the infinite $N_c$, the plateaus of conductance fluctuation
will extend to infinite.

\begin{figure}
\includegraphics[bb=10mm 10mm 205mm 157mm,
width=8.5cm,totalheight=6.2cm, clip=]{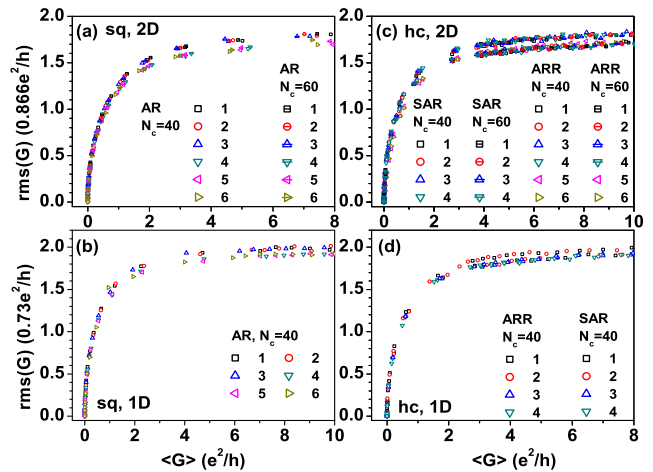} \caption{
(Color online) Corresponding to Fig.\ref{square2D},
Fig.\ref{square1D}, Fig.\ref{honey2D} and Fig.\ref{honey1D}, ${\rm
rms}(G)$ vs small $\langle G\rangle$ ($<10$) are plotted in
panel(a), (b), (c) and (d), respectively.} \label{tail}
\end{figure}

We now take a closer look at each figure discussed above. In the top
panels of Fig.\ref{square2D}-\ref{honey1D}, we can see that all
curves of ${\rm rms}(G)$ vs $\langle G\rangle$ collapse into
universal curves that are slightly separated in the region of
$1<\langle G\rangle<10$. To make the discussion of separate UCF
plateaus quantitative, we plot ${\rm rms}(G)$ vs small $\langle
G\rangle$ ($\langle G\rangle<10$) in Fig.\ref{tail}(a), (b), (c) and
(d) corresponding to Fig.\ref{square2D}, Fig.\ref{square1D},
Fig.\ref{honey2D} and Fig.\ref{honey1D}. In Fig.\ref{tail}, we
clearly see two separate UCF in the regime where $1<\langle
G\rangle<10$. For each UCF plateau, the conductance fluctuation
${\rm rms}(G)$ vs average conductance $\langle G\rangle$ is a
universal function, i.e., it is independent of system parameters
such as $E_F$, $E_0$, $\Delta$, system size and so on and depends
only on $\langle G\rangle$. In fact, not only the ${\rm rms}(G)$
(the second moment), the third, forth, ..., and higher moments are
universal function of $\langle G\rangle$. This means that the
conductance distribution $P(G)$ is a universal function that depends
only on the average conductance $\langle G\rangle$ in addition to
the symmetry and dimensionality of the system.

\begin{figure}
\includegraphics[bb=11mm 9mm 197mm 245mm,
width=6cm,totalheight=7cm, clip=]{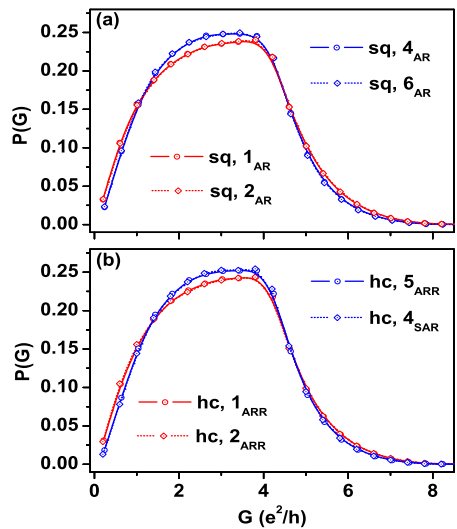} \caption{
(Color online) In the diffusive regime, corresponding to eight
selected parameters with $|E_F|<\Delta$ from Tab.\ref{parameter},
the conductance distribution $P(G)$ obtained from 1,000,000
configurations is plotted for the fixed $\langle G\rangle\simeq3$ in
square [panel (a)] and honeycomb lattices [panel (b)].}
\label{distridiffusi}
\end{figure}
\begin{table}
\caption{In square lattice or honeycomb model, corresponding to
eight selected parameter labeled in Tab.\ref{parameter} with
$|E_F|>\Delta$, the average conductance $\langle G\rangle$ and the
second, third, ..., ninth moments are listed for the first (with
ehD, 1 and 3 column) and the second (ehD broken, 2 and 4 column)
class in the diffusive regime with $\langle G\rangle\simeq3$.}
\begin{tabular}{c c c c c c c c c c}\hline\hline
sq & $\langle G\rangle$ & $\sqrt{\mu_2}$ & $\sqrt[3]{\mu_3}$ &
$\sqrt[4]{\mu_4}$ & $\sqrt[5]{\mu_5}$ & $\sqrt[6]{\mu_6}$ &
$\sqrt[7]{\mu_7}$ & $\sqrt[8]{\mu_8}$ & $\sqrt[9]{\mu_9}$
\\ \hline
$1_{\rm AR}$ & 3.001 & 1.445 & 0.929 & 1.840 & 1.750 & 2.196 & 2.266
& 2.512 & 2.645
\\
$2_{\rm AR}$ & 3.013 & 1.443 & 0.923 & 1.838 & 1.745 & 2.184 & 2.266
& 2.518 & 2.660
\\ \hline
$4_{\rm AR}$ & 2.986 & 1.366 & 0.862 & 1.737 & 1.639 & 2.061 & 2.129
& 2.367 & 2.486
\\
$6_{\rm AR}$ & 2.976 & 1.365 & 0.864 & 1.736 & 1.639 & 2.061 & 2.129
& 2.367 & 2.486 \\
\hline\hline hc & $\langle G\rangle$ & $\sqrt{\mu_2}$ &
$\sqrt[3]{\mu_3}$ & $\sqrt[4]{\mu_4}$ & $\sqrt[5]{\mu_5}$ &
$\sqrt[6]{\mu_6}$ & $\sqrt[7]{\mu_7}$ & $\sqrt[8]{\mu_8}$ &
$\sqrt[9]{\mu_9}$
\\ \hline
$1_{\rm ARR}$ & 2.998 & 1.417 & 0.899 & 1.805 & 1.708 & 2.143 &
2.218 & 2.463 & 2.591
\\
$2_{\rm ARR}$ & 2.988 & 1.421 & 0.901 & 1.809 & 1.713 & 2.148 &
2.224 & 2.470 & 2.602
\\ \hline
$5_{\rm ARR}$ & 3.005 & 1.347 & 0.836 & 1.713 & 1.606 & 2.031 &
2.091 & 2.328 & 2.441
\\
$4_{\rm SAR}$ & 3.009 & 1.344 & 0.833 & 1.710 & 1.602 & 2.027 &
2.087 & 2.324 & 2.438
\\ \hline\hline
\end{tabular}
\label{tabdiffusi}
\end{table}

To demonstrate the conductance distribution has two different
universalities, we plot in Fig.\ref{distridiffusi} the conductance
distribution $P(G)$ obtained from 1,000,000 configurations for a
fixed average conductance $\langle G\rangle\simeq3$ in the square
lattice [panel (a)] and the honeycomb lattice [panel (b)]. In this
figure, we choose eight parameters from Tab.\ref{parameter} with
$|E_F|<\Delta$. We see that for both square and honeycomb lattices,
the conductance distributions corresponding to $E_F=0$ and $E_F\neq
0$ are clearly different. In addition, for each case, $E_F=0$ or
$E_F\neq 0$, conductance distributions for square and honeycomb
lattices are almost the same, as can be seen from
Tab.\ref{tabdiffusi} in which the second, third, ..., ninth moments
are listed for the parameters labeled in Tab.\ref{parameter}
corresponding to the first ($E_F=0$ with ehD) and the second
($E_F\neq 0$ where ehD is broken) classes with fixed $\langle
G\rangle\simeq3$. Here, the n-th moment is defined as
$\mu_n=\left\langle [G-\langle G\rangle]^n\right\rangle$. In
Tab.\ref{tabdiffusi}, the n-th moments labeled by ``$1_{AR}$" and
``$2_{AR}$" correspond to the first class in square lattice, they
are close to the n-th moments labeled by `$1_{ARR}$' and `$2_{ARR}$'
in honeycomb lattice. Since the universal behavior is determined
only by the symmetry and dimensionality, why there are two universal
curves for AR? This can be qualitatively understood as follows.

When the energy of incoming electron is within the superconducting
energy gap, only AR exists. The AR amplitude of total NS system
$T_A$ is contributed by multiple Andreev reflections and can be
expressed in terms of transmission amplitude $t$ and $r$ in the
absence of superconducting leads and the pure AR matrix $r_A$ of the
only NS interface [not consider the clean or disordered normal
scattering region] in the following form\cite{Beenakker}
\begin{eqnarray}
&&T_{A}(\epsilon) = {\rm Tr}[m(\epsilon)m^\dagger(\epsilon)]
\end{eqnarray}
with
\begin{eqnarray}
&&m(\epsilon)= t^e_{12}(\epsilon) M
t^{e,\dagger}_{12}(-\epsilon)\nonumber
\\
&&M =
[I-r^{eh}_A(\epsilon)r^{e,*}_{22}(-\epsilon)r^{eh,T}_A(\epsilon)
r^e_{22}(\epsilon)]^{-1}r^{eh}_A(\epsilon) \label{multi}
\end{eqnarray}
where we have used the electron-hole symmetry relation
$t^h_{21}(\epsilon)=t^{e,*}_{21}(-\epsilon)$,
$r^{he}_A(\epsilon)=r^{eh,T}_A(\epsilon)$ and the symmetry relation
of normal transmission matrix
$t^{e}_{21}(\epsilon)=t^{e,T}_{12}(\epsilon)$ in the absence of
magnetic filed, where `T' denotes transpose. Eq.(\ref{multi}) can be
expanded in power series which gives multiple Andreev reflections.
For qualitative understanding, we can focus on the first term in the
series, i.e., $m(\epsilon)=t_{12}(\epsilon)
t_{12}^\dagger(-\epsilon)$ and $T^{(1)}_{A} =
T_{12}(\epsilon)T_{12}(-\epsilon)$. It is similar for the higher
order of $T_{A}$. Now it is clear why we obtain two universal
conductance distributions for Andreev conductance. For $\epsilon=0$
(with ehD) the total Andreev reflection coefficient $T_{A}$ is
expressed in terms of only one type of normal transmission
coefficient $T(0)$.
For $\epsilon \ne 0$ (ehD
broken), however, $T_{A}$ consists of two kinds of transmission
coefficient $T(\epsilon)$ and $T(-\epsilon)$ that have the
completely different statistics. It is the statistical interference
of $T(\epsilon)$ and $T(-\epsilon)$ that leads to the new universal
conductance distribution.

It should be noted in order to get the uniform statistical
interference, $T(\epsilon)$ and $T(-\epsilon)$ must be separated far
enough from each other, i.e., $\epsilon$ is larger than Thouless
energy. The incident energy $\epsilon$ (related to condensed energy,
equal to $E_F$ in our calculation) is so large that it is comparable
to energy gap $\Delta$, so we must go beyond Andreev approximation
(AA). While in the present works, AA are widely used, it is why the
present works can't present this new symmetry class. We will show
[Fig.\ref{distrilocal}] in the AA, the conductance distribution is
smoothly changed with $E_F$, in stead of the two universal functions
corresponding to $E_F=0$ and $E_F\ne 0$ in the case with non-Andreev
approximation (NAA).

\subsection{Statistical properties in the localized regime}

As we have shown, different universal conductance distributions
corresponding to $E_F=0$ and $E_F\neq0$ are found in the diffusive
regime. It has been demonstrated numerically\cite{ucf4} that the
conductance distribution for a fixed $\langle G\rangle$ in the
localized regime seems to be a universal function which does not
depend on dimensionality (quasi-1D, 2D and quantum dot systems) and
ensemble symmetry (COE, CUE or CSE). For normal-superconductor
hybrid systems, it is interesting to know whether this conclusion is
still valid.

\begin{figure}
\includegraphics[bb=11mm 9mm 201mm 245mm,
width=6cm,totalheight=7cm, clip=]{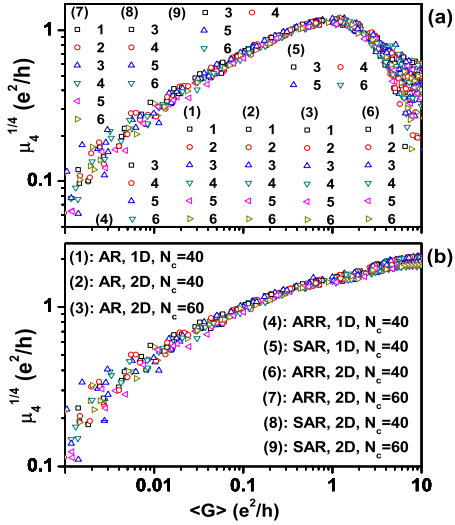} \caption{ (Color
online) the skewness $\gamma_1$ and the kurtosis $\gamma_2$ vs
$\langle G\rangle$ for the 1D square or honeycomb lattice with
$N_c=40$ and 2D square or honeycomb lattice with $N_c=40$ and
$N_c=60$. Different symbols (1)-(9) are described as in panel (b)
and labeled in Tab.\ref{parameter}.} \label{gam}
\end{figure}

\begin{table}
\caption{Same to Tab.\ref{tabdiffusi} except we consider localized
regime with fixed average conductance $\langle G\rangle\simeq0.3$.}
\begin{tabular}{c c c c c c c c c c}\hline\hline
sq & $\langle G\rangle$ & $\sqrt{\mu_2}$ & $\sqrt[3]{\mu_3}$ &
$\sqrt[4]{\mu_4}$ & $\sqrt[5]{\mu_5}$ & $\sqrt[6]{\mu_6}$ &
$\sqrt[7]{\mu_7}$ & $\sqrt[8]{\mu_8}$ & $\sqrt[9]{\mu_9}$
\\ \hline
$1_{\rm AR}$ & .3005 & 0.669 & 0.986 & 1.287 & 1.527 & 1.728 & 1.900
& 2.052 & 2.192
\\
$2_{\rm AR}$ & .2997 & 0.669 & 0.986 & 1.287 & 1.527 & 1.727 & 1.898
& 2.049 & 2.186
\\ \hline
$4_{\rm AR}$ & .3002 & 0.632 & 0.936 & 1.229 & 1.464 & 1.658 & 1.823
& 1.964 & 2.087
\\
$6_{\rm AR}$ & .2990 & 0.631 & 0.936 & 1.229 & 1.464 & 1.660 & 1.826
& 1.970 & 2.099
\\
\hline\hline hc & $\langle G\rangle$ & $\sqrt{\mu_2}$ &
$\sqrt[3]{\mu_3}$ & $\sqrt[4]{\mu_4}$ & $\sqrt[5]{\mu_5}$ &
$\sqrt[6]{\mu_6}$ & $\sqrt[7]{\mu_7}$ & $\sqrt[8]{\mu_8}$ &
$\sqrt[9]{\mu_9}$  \\ \hline $1_{\rm ARR}$ & .2998 & 0.667 & 0.982 &
1.281 & 1.520 & 1.718 & 1.887 & 2.036 & 2.170
\\
$2_{\rm ARR}$ & .3005 & 0.669 & 0.985 & 1.286 & 1.526 & 1.725 &
1.895 & 2.044 & 2.179
\\ \hline
$5_{\rm ARR}$ & .3001 & 0.631 & 0.934 & 1.226 & 1.460 & 1.654 &
1.817 & 1.958 & 2.081
\\
$4_{\rm SAR}$ & .3002 & 0.631 & 0.936 & 1.229 & 1.463 & 1.658 &
1.823 & 1.966 & 2.094
\\
\hline\hline
\end{tabular}
\label{tablocal}
\end{table}

\begin{figure}
\includegraphics[bb=11mm 10mm 202mm 254mm,
width=6cm,totalheight=7cm, clip=]{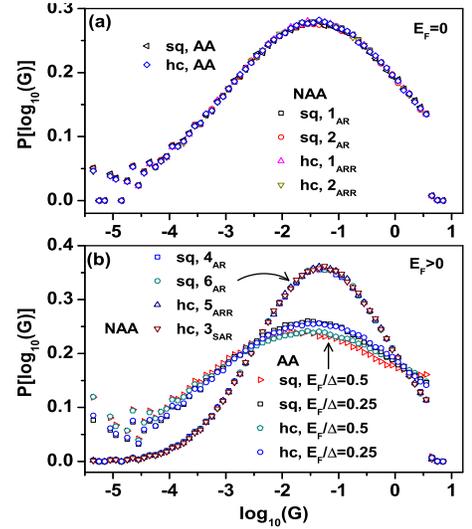}
\caption{(Color online) In the localized regime, corresponding to
selected parameters labeled in Tab.\ref{parameter} with $|E_F|=0$
and $0<|E_F|<\Delta$, the conductance distribution $P[\log_{10}(G)]$
obtained from 1,000,000 configurations are plotted in panel (a) and
panel(b) respectively for the fixed $\langle G\rangle\simeq0.3$ in
square lattice [marked with sq] and honeycomb lattice [marked with
hc]. In addition, we also plot $P[\log_{10}(G)]$ within AA for
$E_F=0$ and $E_F\ne0$ in panel(a) and panel(b), respectively.}
\label{distrilocal}
\end{figure}

There are two ways to examine the universal conductance distribution
$P(G)$: (1). plot $P(G)$ at each $\langle G\rangle$ for different
system parameters to see whether all $P(G)$ collapse into a single
curve. One can only plot $P(G)$ at a few selected $\langle
G\rangle$. (2). plot the moments of $P(G)$ as a function of $\langle
G\rangle$ to see the universal behavior. However one can only plot
several moments of conductance. Here we focus on the higher order
moments $\mu_3$ and $\mu_4$. In Fig.\ref{gam}, we plot
$\sqrt[3]{\mu_3}$ [panel(a)] and $\sqrt[4]{\mu_4}$ [panel(4)] vs
$\langle G\rangle$ for 2D and quasi-1D systems on square and
honeycomb lattices. Symbols (1)-(9) are described as in panel (b)
and labeled in Tab.\ref{parameter}. From the figure, it is clear
that the data do not collapse into a single curve. In this
calculation, we have used only 10,000 configurations per data point
which is not enough to resolve the universality class if any. To
improve this, we fix the average conductance $\langle G\rangle$ and
calculate higher moments by averaging over 1,000,000 configurations.
In Tab.\ref{tablocal}, we choose the same set of parameters as used
in the diffusive regime [Tab.\ref{tabdiffusi}], and tabulate the
average conductance $\langle G\rangle$ and the second, third, ...,
ninth moments for the fixed $\langle G\rangle\simeq0.3$. Similar to
Tab.\ref{tabdiffusi}, two universality classes can be identified.
The first universality class has ehD and consists of data points
from four different set of parameters labeled by ``$1_{AR}$" and
``$2_{AR}$"(square lattice) and labeled by ``$1_{ARR}$" and
``$2_{ARR}$"(honeycomb lattice). The rest of data form the second
universality class where ehD is broken. Hence it is expected that
the conductance distributions for $E_F=0$ (with ehD) and $E_F\neq0$
(without ehD) belong to different universality class in the {\it
localized regime}. This indeed can be seen from
Fig.\ref{distrilocal}(a) and (b) where we have plotted the
conductance distributions of $\log_{10}(G)$ for $E_F=0$ and $E_F\ne
0$. Fig.\ref{distrilocal}(a) shows the conductance distribution with
ehD for six different sets of parameters where two of them are for
AA and the other four are NAA. Obviously, they fall into the same
universality class. In Fig.\ref{distrilocal}(b), we show the data
for the case with broken ehD. We see that four set of data with NAA
collapse into a single curve indicating the universal conductance
distribution that is clearly different from
Fig.\ref{distrilocal}(a). When AA is made, however, the conductance
distribution depends on $E_F$ which is non-universal. The results
from Fig.\ref{distrilocal} show that even in the localized regime,
the Andreev conductance distributions for $E_F=0$ (with ehD) and
$E_F\ne0$ (ehD broken) belong to different universality class.

\subsection{statistics beyond superconducting gap}

\begin{figure}
\includegraphics[bb=12mm 9mm 182mm 176mm,
width=8.5cm,totalheight=8cm, clip=]{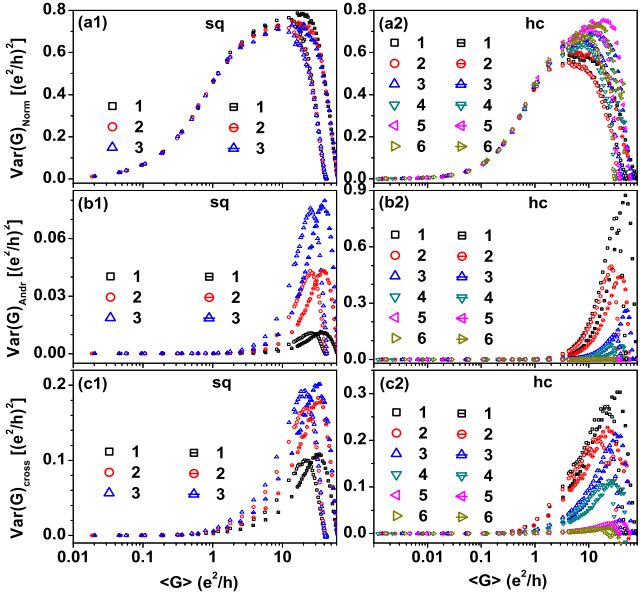}
\caption{ (Color online) ${\rm Var}(G)_{\rm Norm}$, ${\rm
Var}(G)_{\rm Andr}$ and ${\rm Var}(G)_{\rm cross}$, the three
compositions of variance of $G$ vs $\langle G\rangle$ for the 2D
square lattice [the left column] and 2D honeycomb lattice [the right
column] with $N_c=40$ [open symbols] and $N_c=60$ [symbols with
`-'].} \label{NormAndrcross}
\end{figure}

\begin{figure}
\includegraphics[bb=12mm 9mm 203mm 255mm,
width=6.5cm,totalheight=8cm, clip=]{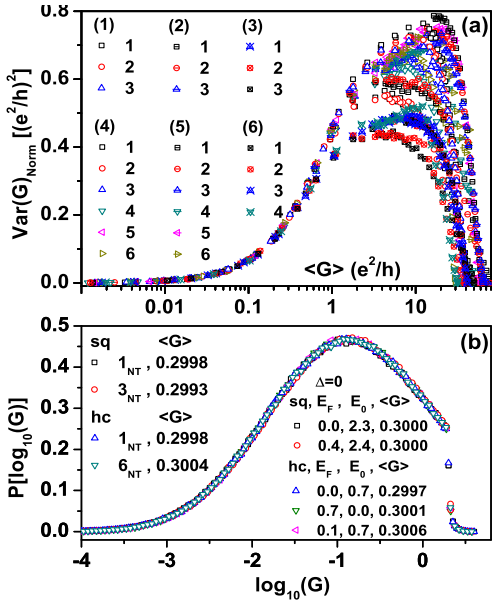} \caption{
(Color online) Panel (a): ${\rm Var}(G)_{\rm Norm}$ vs $\langle
G\rangle$ for the 1D [with $N_c=40$, corresponding to the crossed
symbols] or 2D [with $N_c=40$, corresponding to the open symbols and
$N_c=60$, corresponding to the symbols with `-'] square lattice
[symbols (1), (2) and (3)] or honeycomb lattice [symbols (4), (5)
and (6)]. Panel (b): in a 2D square or honeycomb lattice system with
$N_c=40$, corresponding to four selected parameter with $E_F$ beyond
$\Delta$ and labeled in Tab.\ref{parameter}, conductance
distribution $P[\log_{10}(G)]$ exported from 1,000,000
configurations is plotted. In comparation, we also plot
$P[\log_{10}(G)]$ for different parameters in the normal system with
$\Delta=0$ from 1,000,000 configurations.} \label{Normdistri}
\end{figure}

\begin{table}
\caption{Beyond the superconducting Gap $\Delta$, the average
conductance $\langle G\rangle$ and the second, third, ..., ninth
moments of normal conductance $G_{N}$ are listed in the localized
regime with $\langle G\rangle\simeq0.3$.}
\begin{tabular}{c c c c c c c c c c}\hline\hline
NT & $\langle G\rangle$ & $\sqrt{\mu_2}$ & $\sqrt[3]{\mu_3}$ &
$\sqrt[4]{\mu_4}$ & $\sqrt[5]{\mu_5}$ & $\sqrt[6]{\mu_6}$ &
$\sqrt[7]{\mu_7}$ & $\sqrt[8]{\mu_8}$ & $\sqrt[9]{\mu_9}$
\\ \hline
$1_{\rm sq}$ & .2997 & 0.311 & 0.356 & 0.470 & 0.546 & 0.622 &
0.691 & 0.755 & 0.814 \\
$3_{\rm sq}$ & .2999 & 0.310 & 0.355 & 0.467 & 0.544 & 0.620 &
0.689 & 0.752 & 0.812 \\
$1_{\rm hc}$ & .2998 & 0.310 & 0.357 & 0.472 & 0.551
& 0.631 & 0.703 & 0.771 & 0.835 \\
$6_{\rm hc}$ & .3004 & 0.310 & 0.351 & 0.462 & 0.537 & 0.612 &
0.679 & 0.742 & 0.801 \\
\hline\hline
\end{tabular}
\label{tabnorm}
\end{table}

In previous sub-sections, we have studied the statistical properties
of pure AR assisted conductance with incident energy $|E_F|<\Delta$.
In this sub-section, we will focus on the case in which the incident
energy $E_F$ is above $\Delta$. In this case, conductance is
contributed by both normal transmission and Andreev reflection. The
conductance variance ${\rm Var}(G)$ consists of three terms, the
Andreev conductance fluctuation ${\rm Var}(G)_{\rm Andr}$, the
normal conductance fluctuation ${\rm Var}(G)_{\rm Norm}$ and the
cross term between them ${\rm Var}(G)_{\rm cross}$ [see
Eq.(\ref{conduct}) and Eq.(\ref{Var})]. In Fig.\ref{NormAndrcross},
we plot ${\rm Var}(G)_{\rm Norm}$, ${\rm Var}(G)_{\rm Andr}$ and
${\rm Var}(G)_{\rm cross}$ vs $\langle G\rangle$ for the 2D square
lattice [left panels] and 2D honeycomb lattice [right panels] with
$N_c=40$ [open symbols] and $N_c=60$ [symbols with `-']. Our results
can be summarized as follows. (1) The Andreev related variance ${\rm
Var}(G)_{\rm Andr}$ is drastically suppressed by the disorder. In
localized regime [$\langle G\rangle<1$], due to strong disorder, it
is completely suppressed to almost zero. As a result only ${\rm
Var}(G)_{\rm Norm}$ plays a dominant pole in the localized regime.
(2) in the localized regime, the dominant ${\rm Var}(G)_{\rm Norm}$
exhibits a universal behavior, i.e., it is independent of system
parameters (such as $E_F$, $E_0$, $N_c$, $\Delta$ and so on). In
Fig.\ref{Normdistri}(a), we plot ${\rm Var}(G)_{\rm Norm}$ of 2D
system [Fig.\ref{NormAndrcross}(b1),(b2)] and quasi-1D system for
square lattice and honeycomb lattice. We find that ${\rm
Var}(G)_{\rm Norm}$ in the localized regime is also independent of
dimensionality and type of lattice. It is not surprising since in
localized regime, all AR related process are suppressed by the
strong disorder. In absence of electron-hole conversion, statistics
of NS system are same as that of normal system.

In order to improve the accuracy in the calculation, we also
calculate the higher order moments and conductance distribution by
averaging over 1,000,000 configurations and tabulate average
conductance $\langle G\rangle$ and the second, third, ..., ninth
moments for the fixed $\langle G\rangle\simeq0.3$ in
Tab.\ref{tabnorm}. It is found that the n-th moment for the square
lattice and the honeycomb lattice are the same. Correspondingly, in
Fig.\ref{Normdistri}(b), we plot the conductance distribution of
$\log_{10}(G)$ in a 2D square and honeycomb lattices with $\Delta=0$
and $\Delta\ne0$. The symbols for $\Delta\ne0$ are labeled as in
Tab.\ref{parameter} and the symbols for $\Delta=0$ is described in
Fig.\ref{Normdistri}(b). We see that those data labeled by
``$1_{NT}$" belong to the first class ($E_F=0$), and the other data
belong to the second class ($E_F\neq0$). We can see that when the
incident energy is above the superconducting gap $\Delta$, the
conductance distributions of NS system are almost indistinguishable
from that of normal system with $\Delta=0$. This again confirms that
the normal transmission is dominant, electron-hole conversion and
consequently the ehD is irrelevant in the localized regime.

On experimental side, conductance fluctuation\cite{staley,aris} and
magnetoconductance fluctuation\cite{bra} has been measured for mono
and multi-layer graphene normal systems. The conductance
fluctuations of normal-superconducting hybrid systems (non-graphene)
has also been studied.\cite{hartog} Hence, our results can be
checked experimentally.

\section{conclusion}

Using the tight-binding model, we have carried out a theoretical
study on the sample to sample fluctuation in transport properties of
phase coherent systems with conventional NS hybrid systems or
graphene based NS hybrid systems. Extensive numerical simulations on
quasi-1D or 2D systems show that (1). When $E_F<\Delta$, the UCF due
to AR is found to be roughly doubled comparing to the system in the
absence of the superconducting lead. Denoting the increase factor
$\alpha_0$ through the relation ${\rm rms}(G_{NS})=\alpha_0 ~ {\rm
rms}(G_N)$, we found that the difference between $\alpha_0$ in 2D
system and quasi-1D system is quite large while the difference is
small between Fermi electrons and Dirac electrons. (2). Our results
show that ehD in the NS hybrid system can lead to a new universality
class. In the diffusive regime we found two slightly separated UCF
plateaus, one corresponds to the complete electron-hole symmetry
class (with ehD) and the other to conventional electron-hole
conversion (with ehD broken). In addition, the AR conductance
distribution for the fixed average conductance $\langle G\rangle$ in
diffusive regime also confirms that the new universality class can
be classified using ehD. (3). In the localized regime, we found that
the conductance distribution is a universal function that depends
only on the average conductance and the ehD. We emphasize that one
has to go beyond AA to make sure that the AR conductance
distribution is universal in the localized regime. (4). Finally,
when $E_F$ is beyond $\Delta$, normal transport is present. In
general, the conductance distributions of NS systems and normal
systems are different. In the localized regime, however, the AR is
suppressed significantly by the disorder. Hence in the localized
regime normal transmission dominates the transport processes. In
this case, the ehD is irrelevant and the conductance distribution is
a universal function that depends only on the average conductance in
the localized regime.

$${\bf ACKNOWLEDGMENTS}$$
We gratefully acknowledge the financial support by a RGC grant
(HKU705409P) from the Government of HKSAR.


\begin{references}
\bibitem [*] Electronic address: jianwang@hkusua.hku.hk
\bibitem{book}
B. L. Altshuler, P. A. Lee, and R. A. Webb, Mesoscopic Phenomena in
Solids (North-Holland, Amsterdam, 1991).

\bibitem{Beenakker}
C. W. J. Beenakker, Rev. Mod. Phys. {\bf 69}, 731 (1997).

\bibitem{add}
N.J. Zhu, H. Guo, and R. Harris, Phys. Rev. Lett. {\bf 77}, 1825
(1996); H. U. Baranger and P. A. Mello, Phys. Rev. Lett. {\bf 73},
142 (1994); D. V. Savin, H.-J. Sommers, and W. Wieczorek, Phys. Rev.
B {\bf 77}, 125332 (2008); S. Hemmady, J. Hart, X. Zheng, T. M.
Antonsen, Jr., E. Ott, and S. M. Anlage, Phys. Rev. B {\bf 74},
195326 (2006); Ph. Jacquod and R. S. Whitney, Phys. Rev. B {\bf 73},
195115 (2006).

\bibitem{lee85}
P. A. Lee and A. D. Stone, Phys. Rev. Lett. {\bf 55}, 1622 (1985);
L. B. Altshuler, JETP Lett. {\bf 41}, 648 (1985).


\bibitem{saenz2}
A. Garcia-Martin and J.J. Saenz, Phys. Rev. Lett. {\bf 87}, 116603
(2001); L. S. Froufe-P\'{e}rez, P. Garc\'{l}a-Mochales, P. A.
Serena, P. A. Mello, and J. J. S\'{a}enz, Phys. Rev. Lett. {\bf 89},
246403 (2002).

\bibitem{ucf4}
Z. Qiao, Y. Xing and J. Wang, Phys. Rev. B, {\bf 81}, 085114 (2010).


\bibitem{Beenakker1}
C. W. J. Beenakker, Phys. Rev. B, {\bf 47}, 15763 (1993).

\bibitem{Beenakker2}
P. W. Brouwer and C. W. J. Beenakker, Phys. Rev. B, {\bf 52}, R3868
(1995).

\bibitem{Takane}
Y. Takane and H. Ebisawa, J. Phys. Soc. Jpn. {\bf 61}, 2858 (1992);
J. Bruun, V. C. Hui and C. J. Lambert, Phys. Rev. B, {\bf 49}, 4010
(1994).


\bibitem{randmatri}
A. Altland and Martin R. Zirnbauer, Phys. Rev. B, {\bf 55}, 1142
(1997).

\bibitem{Takane1}
Y. Takane and H. Ebisawa, J. Phys. Soc. Jpn. {\bf 60}, 3130 (1991)

\bibitem{ref16}
H.B. Heersche, P. Jarillo-Herrero, J.B. Oostinga, L.M.K.
Vandersypen, and A.F. Morpurgo, Nature {\bf 446}, 56 (2007).

\bibitem{ref17}
F. Miao, S. Wijeratne, Y. Zhang, U.C. Coskun, W. Bao, and C.N. Lau,
Science {\bf 317}, 1530 (2007).

\bibitem{ref15}
A.F. Andreev, Sov. Phys. JETP {\bf 19}, 1228 (1964).

\bibitem{ref5}
C.W.J. Beenakker, Phys. Rev. Lett. {\bf 97}, 067007 (2006).

\bibitem{Wangbg}
Q. Zhang, D. Fu, B. Wang, R. Zhang, and D. Y. Xing, Phys. Rev. Lett.
{\bf 101}, 047005 (2008).

\bibitem{Xing}
Y.X. Xing, J. Wang and Q.F. Sun, unpublished.

\bibitem{ehsymm}
P. Jarillo-Herrero, S. Sapmaz, C. Dekker, L. P. Kouwenhoven H. S. J.
van der Zant, Nature, {\bf 429}, 389 (2004).

\bibitem{ehsymm1}
H. Pan, T.-H. Lin, and D. Yu, Phys. Rev. B {\bf 70}, 245412 (2004).

\bibitem{Landauer}
R. Landauer, IBM. J. Res. Dev. {\bf 1}, 223 (1957); ibid., Z. Phys.
B {\bf 68}, 217 (1987).

\bibitem{Imry}
{\sl Directions in Condensed Matter Physics}, edited by G. Grinstein
and G. Mazenko (world Scientific, Singapore), p.101.

\bibitem{Buttiker}
M. B\"{u}ttiker, Phys. Rev. B {\bf 57}, 1761 (1988).

\bibitem{ref19}
L. Sheng, D. N. Sheng, and C. S. Ting, Phys. Rev. Lett. {\bf 94},
016602 (2005); L. Sheng, D. N. Sheng, C. S. Ting, and F. D. M.
Haldane, Phys. Rev. Lett. {\bf 95}, 136602 (2005).

\bibitem{ref20}
D. N. Sheng, L. Sheng, and Z. Y. Weng, Phys. Rev. B {\bf 73}, 233406
(2006); W. Long, Q.-F. Sun, and J. Wang, Phys. Rev. Lett. {\bf 101},
166806 (2008).

\bibitem{Nambu}
Y. Nambu, Phys. Rev. {\bf 117}, 648 (1960).

\bibitem{realGap}
P. G. de Gennes, {\sl Superconductivity of Metals and Alloys}
Benjamin, New York (1996).

\bibitem{footnote}
For the electron-hole symmetry, we can calculate only the current
$J_e$ contributed by the electron and double it to get total
current.

\bibitem{transfer}
D. H. Lee and J. D. Joannopoulos, Phys. Rev. B {\bf 23}, 4997
(1981); {\sl ibid}, {\bf 23}, 4988 (1981).


\bibitem{LiDafang}
D. Li and J. Shi, Phys. Rev. B {\bf 79}, 241303(R) (2009).

\bibitem{shape2D}
A. Rycerz, J. Tworzyd{\l}o and C. W. J. Beenakker, Eur. Phys. Lett.
{\bf 79}, 57003 (2007).

\bibitem{staley}
N.E. Staley, C.P. Puls, and Y. Liu, Phys. Rev. B {\bf 77}, 155429
(2008).

\bibitem{aris}
C. Ujeda-Aristizabal, M. Monteverde, R. Weil, M. Ferrier, S. Gueron,
and H. Bouchiat, Phys. Rev. Lett. {\bf 104}, 186802 (2010).

\bibitem{bra}
S. Branchaud, A. Kam, P. Zawadzki, F.M. Peeters, and A.S. Sachrajda,
Phys. Rev. B {\bf 81}, 121406 (2010).

\bibitem{hartog}
S.G. den Hartog, C.M.A. Kapteyn, B.J. van Wees, and T.M. Klapwijk,
Phys. Rev. Lett. {\bf 76}, 4592 (1996).

\end{references}
\end{document}